\documentclass[11pt]{article}
\usepackage{amsmath,amsthm,amssymb,amscd}
\usepackage{xurl}
\usepackage{hyperref}
\usepackage{geometry}
\usepackage{tikz-cd}
\usepackage{booktabs} 
\usepackage{graphicx}
\usepackage{multirow}
\usepackage{tikz}
\usetikzlibrary{calc,positioning,3d}
\usepackage[normalem]{ulem} 
\usepackage[justification=raggedright,singlelinecheck=false]{caption}
\usepackage{imakeidx}
\makeindex
\usepackage{booktabs}
\usepackage{siunitx}
\usepackage{graphicx}
\usepackage[font=small,labelfont=normalfont]{caption}
\usepackage{booktabs,tabularx}
\newcolumntype{Y}{>{\raggedright\arraybackslash}X}

\usepackage{float}

\usepackage{hyperref}
\usepackage{xr-hyper}
\usepackage{caption}
\usepackage{subcaption}

\usepackage{makecell} 

\theoremstyle{definition}

\title{Commutative Algebra Modeling in Materials Science  -- A Case Study on Metal-Organic Frameworks (MOFs)}
\author{
	Caleb Simiyu Khaemba$^{~1}$,
	Hongsong Feng$^{`2}$, Dong Chen$^{~1}$,  \\	
    Chun-Long Chen \footnote{Corresponding author: Chunlong Chen (chunlong.chen@pnnl.gov).} $^3$
		and Guo-Wei Wei\footnote{Corresponding author: Guo-Wei Wei (weig@msu.edu).~}$^{~1,4,5}$ \\		
			$^1$Department of Mathematics,\\
		Michigan State University, MI 48824, USA.\\
		$^2$Department of Mathematics and Statistics,\\
		University of North Carolina at Charlotte, Charlotte, NC 28223, USA\\
        $^3$Physical Sciences Division, Pacific Northwest National Laboratory, Richland,\\ Washington 99354, USA \\ 
		$^4$Department of Electrical and Computer Engineering,\\
		Michigan State University, MI 48824, USA.\\
		$^5$Department of Biochemistry and Molecular Biology,\\
		Michigan State University, MI 48824, USA.
	}
	
\begin{document}
 \maketitle 

\begin{abstract}

 Metal-organic frameworks (MOFs) are a class of important crystalline and highly porous materials whose hierarchical geometry and chemistry hinder interpretable predictions in materials properties. Commutative algebra is a branch of abstract algebra that has been rarely applied in data and material sciences. We introduce the first ever commutative algebra modeling and prediction in materials science. Specifically, category-specific commutative algebra (CSCA) is proposed as a new framework for MOF representation and learning. It integrates element-based categorization with multiscale algebraic invariants to encode both local coordination motifs and global network organization of MOFs.  These algebraically consistent, chemically aware representations enable compact, interpretable, and data efficient modeling of MOF properties such as Henry's constants and uptake capacities for common gases.  Compared to traditional geometric and graph-based approaches, CSCA achieves comparable or superior predictive accuracy while substantially improving interpretability and stability across data sets. By aligning commutative algebra with the chemical hierarchy, the CSCA establishes a rigorous and generalizable paradigm for understanding structure and property relationships in porous materials and provides a nonlinear algebra-based framework for data-driven material discovery.

\end{abstract}
%

\clearpage
\section{Introduction}
\label{sec:introduction}

 Materials science concerns material's structure, properties, processing, and performance. A special class of materials is metal-organic frameworks (MOFs), crystalline networks with open porosity and tunable chemistry. MOFs are attractive for adsorption, gas storage, separations, catalysis, sensing, and transport~\cite{sharp2021nanoconfinement, senkovska2025adsorption, zulfiqar2025metal}. They are assembled from metal nodes and organic linkers, and the way these building blocks connect creates a large design space with controllable ring sizes, pore shapes, and network connectivity that strongly influence material properties~\cite{o2008reticular, ji2020pore, wilmer2012large}. This flexibility can be further enhanced by varying the choice of nodes and linkers or by applying post-synthetic treatments~\cite{aunan2021modulation, keshavarz2023dissecting}, which has enabled applications ranging from energy conversion and storage to drug delivery~\cite{wang2013metal}. One important example is polyoxometalate-based MOFs (POM-MOFs), which combine broad networks with cage-like metal oxygen clusters such that global pore connectivity and local site chemistry interact~\cite{chen2023polyoxometalate}. Important adsorption and transport properties such as Henry's constants for oxygen and nitrogen, oxygen and nitrogen uptakes, and self diffusivities depend on both the chemistry of the cluster environment and the architecture of the pore network~\cite{chen2025interaction}. Recent multiscale studies show that prediction of these adsorption and transport properties improves when chemical information is combined with geometric and topological features of the pore network~\cite{chen2025interaction}, motivating our approach of modeling MOFs by linking chemical environments at the local scale with framework wide structural characteristics.

Traditional experimental characterization remains the benchmark, but scales poorly with the design space. Computational workflows broaden coverage: grand canonical Monte Carlo (GCMC) often with Widom insertion for Henry's constant and uptake, classical molecular dynamics (MD) for diffusion, and density functional theory (DFT) for charge assignment and adsorption energetics~
\cite{kohn1996density,ColonSnurr2014, sholl2016seven}.
Still, sweeping conditions with GCMC, running long MD trajectories, and repeating DFT over thousands of frameworks are computationally intensive. To increase throughput, data-driven surrogates compress each framework into fixed-length features and learn a mapping to targets; for example, framework guest energy histograms with sparse regression enable rapid screening validated by simulation and experiment~\cite{bucior2019energy}.

Supervised machine learning (ML) benefits from standardized structures and labels in CoRE MOF, hMOF, and QMOF~\cite{chung2014computation,wilmer2012large,rosen2021machine}. Chemistry-informed priors such as HSAB (hard-soft acid-base) labels for metals and linkers, and compatibility scores can improve generalization and interpretability for stability and adsorption~\cite{han2025development}. Deep learning (DL) models such as convolutional neural networks (CNNs) learn directly from crystal graphs, bypassing hand crafted descriptors and linking predictions to local chemical environments~\cite{xie2018crystal}. Pretrained multi-modal transformers that fuse atom graph and energy grid embeddings can be fine-tuned to predict adsorption, diffusion, and electronic properties with transferable, attention-based insight~\cite{kang2023multi}. Although recent methods have captured rich MOF structures, they still struggle in giving accurate predictions for some properties. Descriptor-based machine learning underperforms for certain key MOF properties, even with energy histogram features, for example when capturing the microenvironment of metal clusters and their interactions with substrates. Meanwhile, DL approaches (CNNs and Transformers) often entail high computational cost and substantial data requirements, constraining their utility in small data regimes~\cite{xie2018crystal,kang2023multi}. Complementary directions include feature driven pipelines using energy based surrogates, and geometric and chemical summaries such as energy histograms~\cite{bucior2019energy}, and representation learning with graph neural networks and MOF-specific Transformers that derive features directly from structure~\cite{kang2023multi,PMTransformer2023}. Most MOF ML studies still rely on conventional descriptors such as geometry and atom types. Recent literature, however, has highlighted topology-based alternatives that capture network connectivity and periodicity~\cite{glasby2024topological,jiang2021topological,chen2025category}. Several works explore mathematically grounded constructions from algebraic topology and spectral graph theory such as topological Laplacians and persistent Laplacians for robust, high-level representations~\cite{nguyen2020review,nguyen2019agl,wang2020persistent}. However, these approaches mainly capture connectivity through spectra and persistence, and they do not encode the underlying algebraic structure or its evolution across scales.

A natural next step is to explore more fundamental and interpretable mathematical frameworks that can represent MOF structures at multiple scales. Commutative algebra, which studies commutative rings, ideals, and modules, provides such a foundation by offering algebraic tools to encode geometry and connectivity~\cite{miller2005graduate,bruns1998cohen}. Its use in data science and AI is still emerging. Persistent Stanley-Reisner theory (PSRT) bridges commutative algebra, algebraic topology, and machine learning~\cite{suwayyid2025persistent}. Classical Stanley-Reisner theory encodes a simplicial complex as a square free monomial ideal~\cite{francisco2014survey,ha2008monomial}; PSRT tracks this structure across multiple geometric scales to yield computable, interpretable representations, such as persistent graded Betti numbers (via Hochster’s formula), persistent $f$-vectors and $h$-vectors, and persistent facet ideals with facet barcodes summarizing births and deaths~\cite{suwayyid2025persistent}. These multiscale features enable the commutative-algebra analysis of point clouds and have found great success in predicting protein-ligand binding \cite{feng2025caml}, protein-nucleic acid binding \cite{zia2025cap, zia2025gbnl},  disease and mutation association \cite{wee2025commutative}, genomic and phylogenetic analysis\cite{suwayyid2025cakl}. 
A comparative study has been carried out on the interpretability and representability of commutative algebra, algebraic topology, and topological spectral theory for real-world data \cite{ren2025interpretability}. 
In general,   commutative algebra approaches complement topological data analysis (TDA) and topological deep learning (TDL)~\cite{su2025topological,wee2025review,papamarkou2024position}.

In this work, we present the first ever commutative algebra model in  materials science. We propose category-specific commutative algebra (CSCA) for the prediction of MOF properties such as gas adsorption. Our method is benchmarked against three large-scale baselines: MOFTransformer, PMTransformer, and descriptor-based models~\cite{kang2023multi,park2023enhancing,orhan2021prediction}. The CSCA framework encodes category-wise information about metal nodes and organic linkers into compact and interpretable features, which serve as inputs to a simple supervised learner. Our goal is to achieve higher accuracy on Henry’s constants and uptakes for $\mathrm{N}_2$ and $\mathrm{O}_2$, while lowering computational cost compared to deep learning models and improving interpretability by tracing the predictions back to specific chemical categories.

The rest of this work is organized as follows. Section \ref{sec:Results} is devoted to the result and discussion. The proposed method is described in Section \ref{sec:Methods}. This paper ends with a conclusion. 

\section{Result and Discussion}
\label{sec:Results}
\subsection{Workflow}
\begin{figure}[H]
    \centering
    \fbox{\includegraphics[width=1\linewidth]{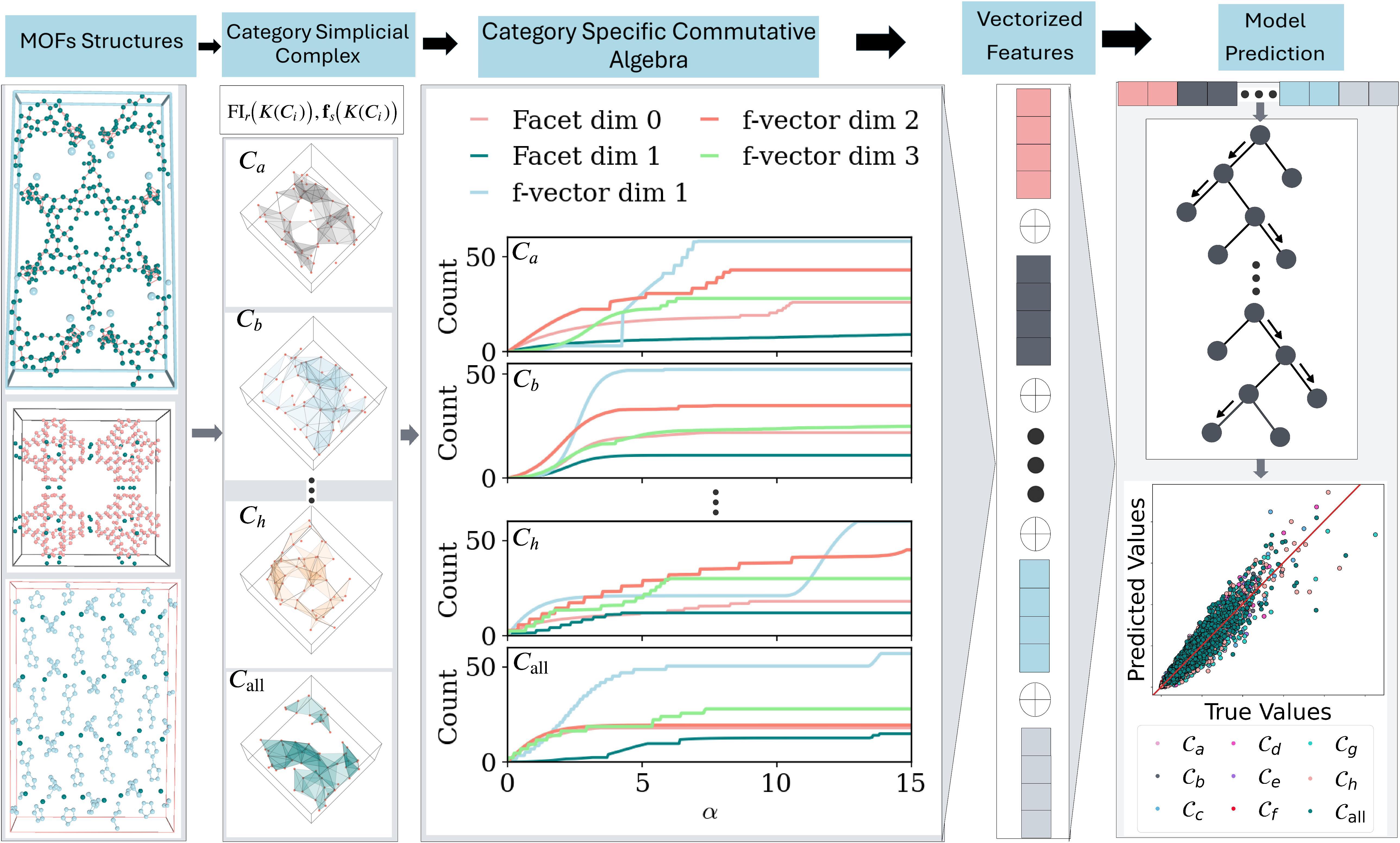}}
    \caption{Category-Specific Commutative Algebra (CSCA) pipeline for MOF property prediction. (a) Starting point: each MOF structure is uniformly rescaled to a cubic supercell of side $64\,\text{\AA}$. (b) Category simplicial complex construction: for each atom group $C_i \in \{C_a,\dots,C_h,C_{\text{all}}\}$, atoms of that category are extracted and used to form  the simplicial complex $K(C_i)$ using the $\alpha$-complex filtration. (c) Algebraic curve generation: along the filtration parameter $\alpha$, we compute two types of descriptors: facet interval curves $FI_r(K(C_i);\alpha)$ for dimensions $r=0,1$ (capturing connected components and loops) and $f$-vector curves $f_s(K(C_i);\alpha)$ for dimensions $s=1,2,3$ (capturing counts of edges, triangles, and tetrahedra). 
    (d) Feature vectorization: summary statistics of these curves are concatenated across all categories and dimensions into a unified feature representation. (e) Model prediction: the resulting feature vectors are used as input to a gradient boosting model, which predicts adsorption and transport properties of MOFs. This stepwise framework connects MOF structure to predictive machine learning via CSCA.}
    \label{fig:flowchat}

\end{figure}

The key steps of the category-specific commutative algebra model are outlined in an end to end pipeline as presented in Figure \ref{fig:flowchat}. Each MOF is first rescaled to a uniform $64\,\text{\AA}\times 64\,\text{\AA}\times 64\,\text{\AA}$ supercell so that all structures share a common length scale. We then partition the atoms into element categories namely $C_a,\ldots,C_h, C_{\mathrm{all}}$, where $C_{\mathrm{all}}$ represents the full set, see Table \ref{tab:4}. Common backbone species such as C, H, O, and N are separated, while chemically similar and less frequent metals are aggregated. This preserves distinct chemical roles and mitigates sparsity in category-wise statistics. We build a simplicial representation for the atoms in each category $C_a,\ldots,C_h, C_{\mathrm{all}}$. A $k$-simplex is the set determined by $k{+}1$ vertices. The $0$-simplices are atoms or vertices, $1$-simplices are bonds or edges, $2$-simplices are triangles, and $3$-simplices are tetrahedra. For each category $C_i\in\{C_a,\ldots,C_h,C_{\text{all}}\}$, we construct the $\alpha$-complex on its atoms and take the downward closure to obtain the geometric simplicial complex $K(C_i)$; see Fig.~\ref{filtration}. With simplices included whenever their circumsphere radius is at most $\alpha$, we build features from a radius filtration up to $12\,\text{\AA}$. We record $f$-vector entries for edges, triangles, and tetrahedra and facet counts of dimensions $0$ and $1$ for all categories, $C_a,\ldots,C_h$, and $C_{\mathrm{all}}$. The $f$-vector describes how many bonds, surfaces, and $3$-D cages exist at each radius $\alpha$, while facet intervals measure the lifetimes of atoms and bonds before they are filled in by higher-dimensional structures. They are uniformly sampled throughout $\alpha$  and concatenated over $C_a,\ldots,C_h$, and $C_{\mathrm{all}}$ to produce fixed-length vectors. Features designed based on these curves and their statistics are fed into gradient-boost decision tree (GBDT) to build models.  

\begin{figure}[H]
    \centering
    \fbox{\includegraphics[width=1\linewidth]{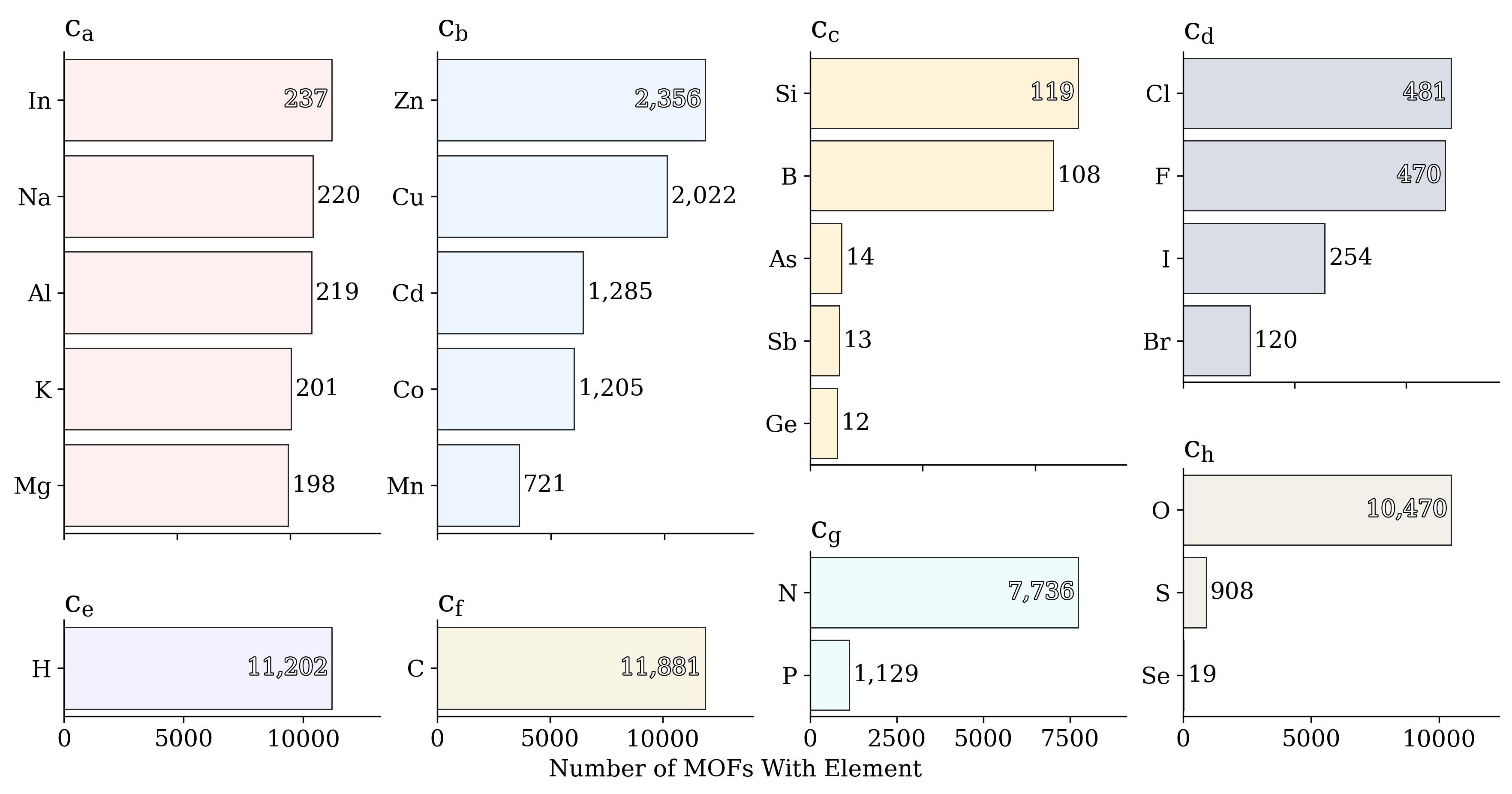}}
   \caption{ Top five most frequent elements within each compositional category 
    $C_a, \dots, C_h$ in the CoRE MOF 2019 dataset. Each bar represents the number of frameworks containing a given element. The distribution reveals prevalent elements in each chemical group, such as alkali and alkaline earth metals in $C_a$, transition metals in $C_b$, and light nonmetals (C, N, O, H) in $C_e, \dots, C_h$. These trends form the compositional backbone of the CSCA representation, linking chemical diversity to category-specific algebraic features used for adsorption property prediction.
    }

    \label{fig:elements}
\end{figure}

Figure \ref{fig:elements} shows the elemental distribution across the dataset, which serves as the basis for defining our categories. For each category $C_a, \dots, C_h, C_{\mathrm{all}}$, Figure \ref{fig:elements} reports the five most frequently occurring components ordering them according to how often they appear in the MOF dataset. Carbon in $C_f$, hydrogen in $C_e$, and oxygen in $C_h$ play significant roles in organic linkers and inorganic nodes, accounting for over $10{,}000$ of MOFs. The relative abundance of nitrogen and phosphorus in $C_g$ supports functional diversity in linkers. Zn and Cu are among the thousands of transition metals in $C_b$, indicating their significance as secondary building blocks. In contrast, the halogens in $C_d$ and elements in $C_c$ (such as Si, B, and As) only occur in the hundreds, indicating niche functional functions as opposed to backbone constituents. Overall, Figure \ref{fig:elements} depicts the long tail contribution of rarer atoms that may have a significant impact on material characteristics as well as the dominance of core elements.

\subsection{Dataset Overview}
\begin{figure}[H]
    \centering
    \fbox{\includegraphics[width=0.88\linewidth]{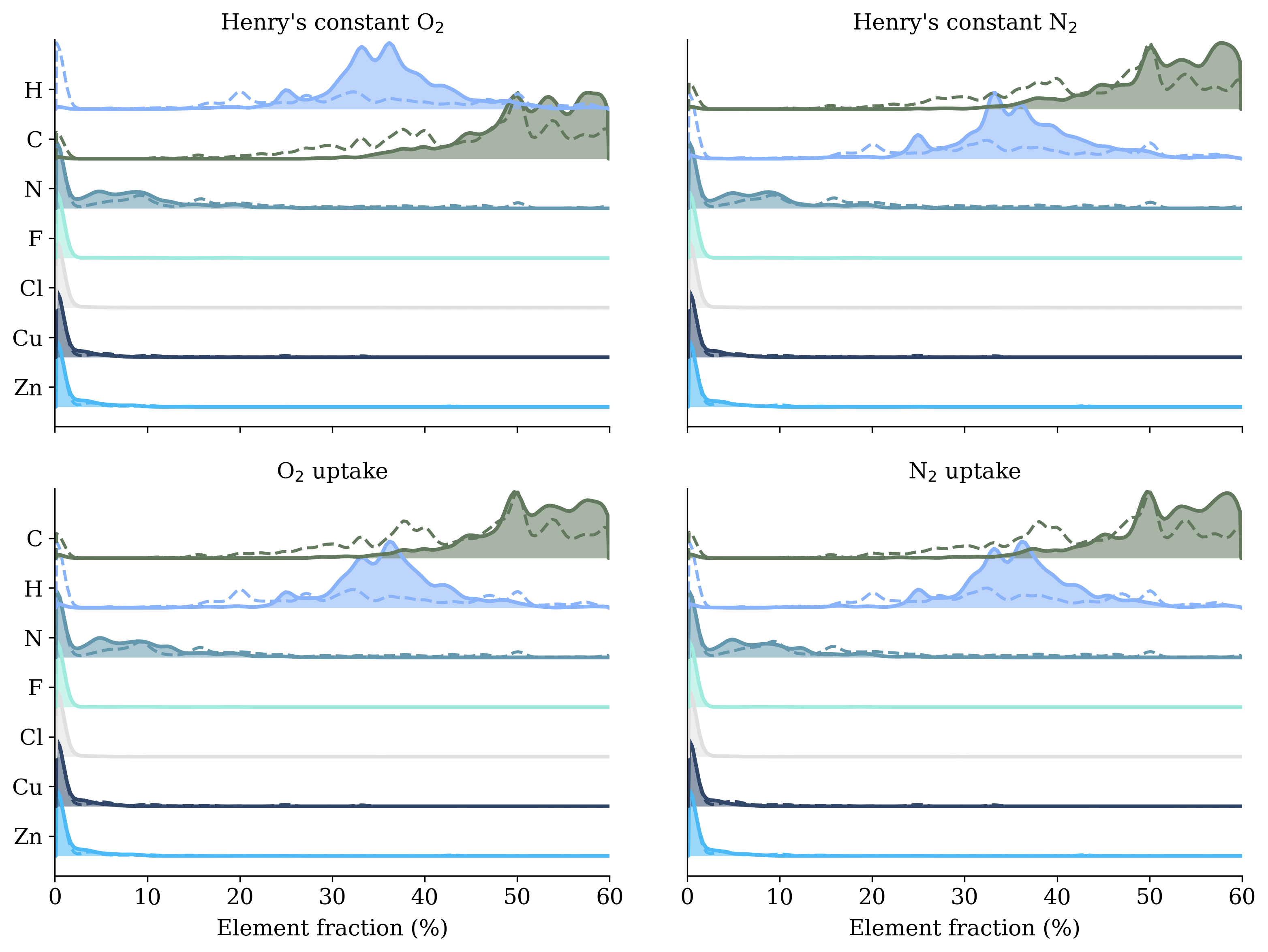}}
    \caption{Element fraction ridgeline plots across four MOF property datasets. 
    For each element, two distributions are shown: filled line $=$ top $25\%$ (MOFs with high property values) and dashed line $=$ bottom $25\%$ (MOFs with low property values). A rightward shift of the filled curve indicates enrichment in high-value MOFs. Observed trends: C and H are enriched; N shows mild enrichment; F, Cl, Cu, and Zn exhibit little or no shift.}

    \label{fig:ridge}
\end{figure}

Our analysis centers on four property datasets: the Henry's constants of oxygen and nitrogen, as well as their uptake capacities. The only property columns in each dataset are O$_2$ uptake (mol\,kg\(^{-1}\)), N$_2$ uptake (mol\,kg\(^{-1}\)), Henry’s constant for O$_2$, and Henry’s constant for N$_{2}$. The element fraction ridgelines for these four property dataset is shown in Figure \ref{fig:ridge}. For each property in each dataset, we compute the 25th and 75th percentiles  and define two groups: Low for MOFs at bottom 25\% and High for MOFs at top 25\%. The middle 50\% is not used in the ridge-line contrast. We then keep rows with a valid numeric property and convert the element count columns (H, C, N, O, metals, halogens) to the fraction of atoms in that MOF; this fraction is the number of atoms of that element divided by the total atom count, reported as a percentage. We draw two ridge-lines for each element: filled line for the High group and dashed line for the Low group. 

A right shift of the filled curve means that the element is more abundant in MOFs with higher property values; a left shift means less; a strong overlap means little or no association. Quartiles provide two equal-sized groups, are robust to outliers, and avoid assuming a linear relationship. This lets us determine which atoms are associated with better performance for those properties.  In all four panels, C and H shift right, N shows a small right shift, and F, Cl, Cu, and Zn stay near zero with substantial overlap.

\subsection{Category-Specific Model Development in a Commutative Algebra Structure}\label{subsec:model dev}
 We present category-specific topological representations for MOFs in this study. Table~\ref{tab:4} outlines the classification of atoms into categories \(C_a,\dots,C_h\) with additional category, \(C_{\text{all}}\), which contains all atoms. We create an alpha complex and perform its filtration, capped at \(12\)\text{\AA}\, for every MOF and category (including \(C_{\text{all}}\) ). To describe higher order characteristics, we consider two sets of features. First, we compute $f$-vector components in dimension $1,2,$ and $3$ correspoding to the counts of edges, triangles, and tetrahedra. Second, we construct facet ideals in dimension $0$ and $1$, which capture algebraic relations among vertices and edges. Together, these features summarize both the combinatorial and algebraic aspects of the underlying structure. We also calculate summary statistics: mean, variance, minimum, and maximum of the corresponding filtration values for every category and simplex dimension to guarantee robustness and comparability. A single embedding vector that captures global organization together with category-specific commutative algebra invariants is produced by concatenating the facet ideal features, $f$-vectors features, and summary statistics across categories. This vector is then appropriate for further learning and analysis.

\subsection{Predicting MOF Properties from Category-Specific Commutative Algebra Features}
In this study, four MOF properties are predicted to assess our category-specific commutative algebra (CSCA) model: uptake capacities for \(\mathrm{N}_2\) and \(\mathrm{O}_2\) \((\text{mol}\,\text{kg}^{-1}\,\text{Pa}^{-1})\) and Henry's constants for \(\mathrm{N}_2\) and \(\mathrm{O}_2\) \((\text{mol}\,\text{kg}^{-1})\).  Table~\ref{tab:2} and Section~\ref{dataset} provide a detailed explanation of the data sets and their preparation.
\begin{table}[ht!]
\caption{Presents a comparison of CSCA performance with existing models across multiple MOF datasets.}
\centering
\label{tab:1}
\renewcommand{\arraystretch}{1.7}
\setlength{\tabcolsep}{8pt}
\setlength{\arrayrulewidth}{0.6pt} 
\begin{tabular}{|llccc|} 
\hline
\textbf{Dataset} & \textbf{Method} & $R^2$ & MAE & RMSE \\
\hline
\multirow{4}{*}{Henry’s constant N$_2$} 
 & \textbf{CSCA}        & $0.78$ & $5.30\times10^{-7}$ & $7.74\times10^{-7}$ \\
 & Descriptor-based\cite{orhan2021prediction}& 0.70 & --                  & $8.94\times10^{-7}$ \\
 & MOFTransformer\cite{kang2023multi}  & --   & -- & -- \\
 & PMTransformer\cite{park2023enhancing}   & --  & -- & -- \\
\hline
\multirow{4}{*}{Henry’s constant O$_2$} 
 & \textbf{CSCA }        & $0.81$ &$5.29\times10^{-7}$  &  $8.07\times10^{-7}$\\
 & Descriptor-based  \cite{orhan2021prediction}          & 0.74 & --                 & $9.60\times10^{-7}$ \\
 & MOFTransformer   \cite{kang2023multi}           & --   & --& -- \\
 & PMTransformer    \cite{park2023enhancing}           & --   & -- & -- \\
\hline
\multirow{4}{*}{N$_2$ uptake (mol\,kg$^{-1}$)} 
 & \textbf{CSCA }        & $0.78$ &$5.16\times10^{-2}$  & $7.62\times10^{-2}$\\
 & Descriptor-based  \cite{orhan2021prediction}          & 0.71 & --                  & $8.62\times10^{-2}$ \\
 & MOFTransformer    \cite{kang2023multi}          & $0.78$ & $7.10\times10^{-2}$ & -- \\
 & PMTransformer   \cite{park2023enhancing}            & --    & $6.90\times10^{-2}$ & -- \\
\hline
\multirow{4}{*}{O$_2$ uptake (mol\,kg$^{-1}$)} 
 & \textbf{CSCA }        & $0.84$ & $4.79\times10^{-2}$&$7.25\times10^{-2}$\\
 & Descriptor-based  \cite{orhan2021prediction}          & $0.74$& --                  & $9.28\times10^{-2}$ \\
 & MOFTransformer   \cite{kang2023multi}           & $0.83$ & $5.10\times10^{-2}$ & -- \\
 & PMTransformer  \cite{park2023enhancing}             & --    & $5.30\times10^{-2}$ & -- \\
\hline
\end{tabular}
\end{table}

\begin{figure}[H]
    \centering
    \includegraphics[width=0.85\linewidth]{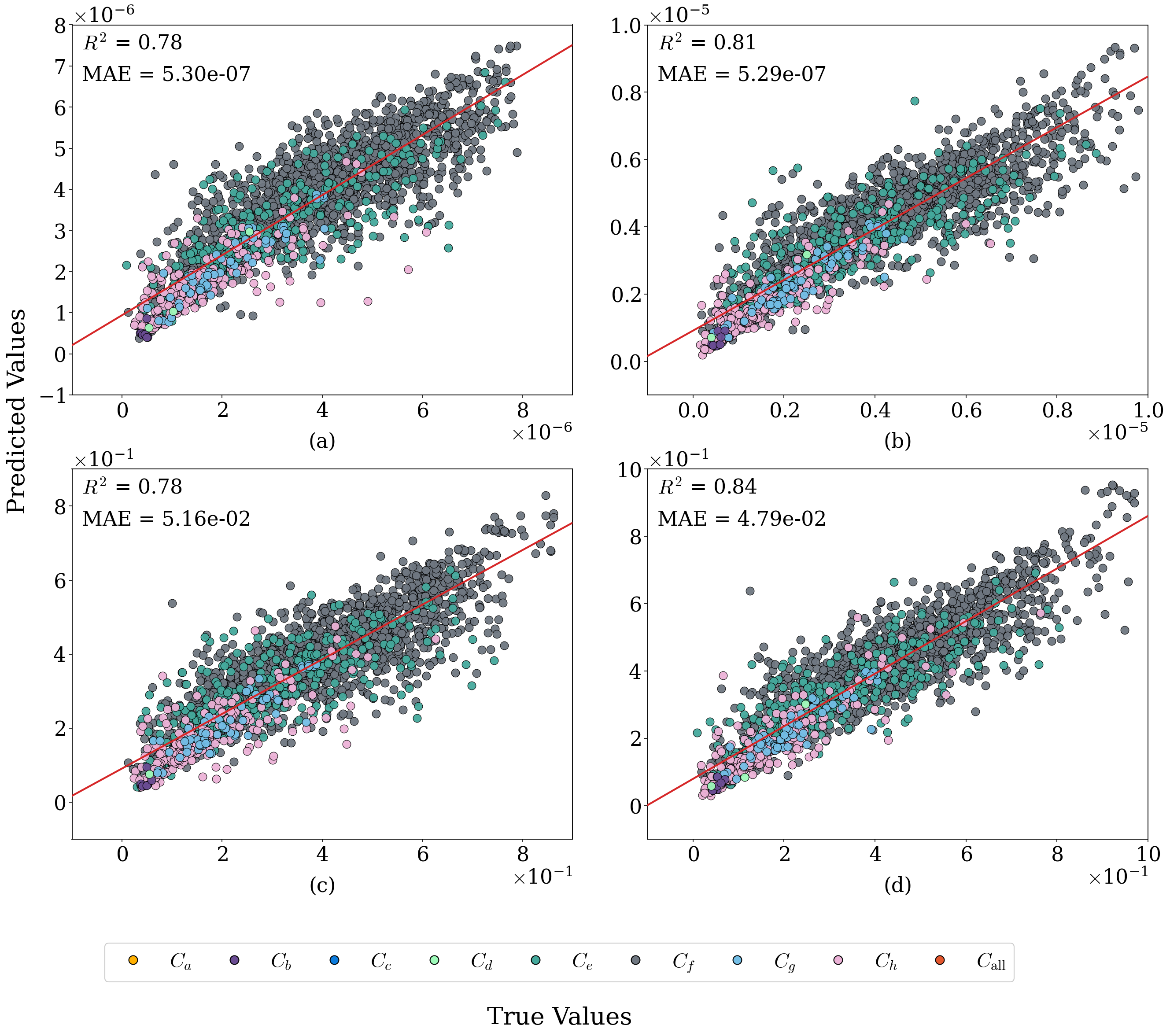}
    \caption{Predicted versus true values for the four property datasets: 
    (a) Henry's constants for $\mathrm{N}_2$; 
    (b) Henry's constants for $\mathrm{O}_2$; 
    (c) uptake capacities for $\mathrm{N}_2$; and 
    (d) uptake capacities for $\mathrm{O}_2$.
    Each point represents a single MOF sample, where the predicted value $\hat{y}_i$
    is obtained from the trained category-specific commutative algebra (CSCA) model
    using the corresponding feature vector of that sample. Points are colored according to their dominant atomic category 
    ($C_a,\dots, C_h$ and $C_{\mathrm{all}}$), determined from atomic composition
    as defined in Table~\ref{tab:3}. All MOF samples are predicted by the same model; colors therefore indicate the primary chemical group to which each sample belongs rather than different feature types or separate models. Each panel reports two evaluation metrics in the upper-left corner: the mean absolute error (MAE) and the coefficient of determination ($R^2$), where $R^2$ measures the proportion of variance in the true values $y_i$ explained by the predicted values $\hat{y}_i$.}

    \label{fig:scatter}
\end{figure}

The performance of our model on the four property datasets is summarized in Figure~\ref{fig:scatter}, and the performance of our model compared to other models is shown in Table~\ref{tab:1}. The prediction findings in Figure \ref{fig:scatter} show good agreement between the predicted and true values in the four data sets when training, validating and testing using a 80{:}10{:}10 split. For each dataset, we average the results of $100$ fits ($10$ random data splits, each trained with $10$ independent  seeds) of the model run with various seeds. The accuracy and stability of the model are highlighted by the mean $R^2$ and MAE displayed in the upper left corner of each panel. Our category-specific commutative algebra model is benchmarked by comparing its performance to three large-scale baseline models: MOFTransformer, PMTransformer and Descriptor-based\cite{kang2023multi,park2023enhancing,orhan2021prediction}, where available. The category specific  commutative  algebra model utilizes a single universal set of hyperparameters presented in Table \ref{tab:4} across all datasets without validation-tuned settings, consistently outperforms descriptor baselines and competitive with or better than the transformer techniques as summarized in Table \ref{tab:1}.

Color-coded categories demonstrate a significant class imbalance, with $C_f$ accounting for the vast majority of samples, as seen in Figure~\ref{fig:scatter}. 
 In all four datasets, there is no discernible class division; instead, the categories merge into a single overlapping diagonal band. Although minority categories  emerge as thinner overlays that reveal somewhat larger dispersion, particularly in mid to high ranges, $C_f$ generally sustains the dense core along the regression line. Similar scaling behavior is shown at the extremes, where all categories taper but stay in line with the diagonal. The abundance of $C_f$ dominates the global metrics $R^2$ and MAE, but there is no indication of category-specific bias because every category exhibits the same increasing trend. However, there is a slight variation in accuracy; $C_g$ and $C_b$ cluster more closely around the regression line, whereas $C_h$ and $C_e$ shows higher dispersion at lower to mid values. Overall, this suggests that the high global performance represents consistent predicted accuracy across chemistry rather than being only a result of the majority class.

\subsection{Feature Construction}
\begin{figure}[H]
  \includegraphics[width=1\linewidth]{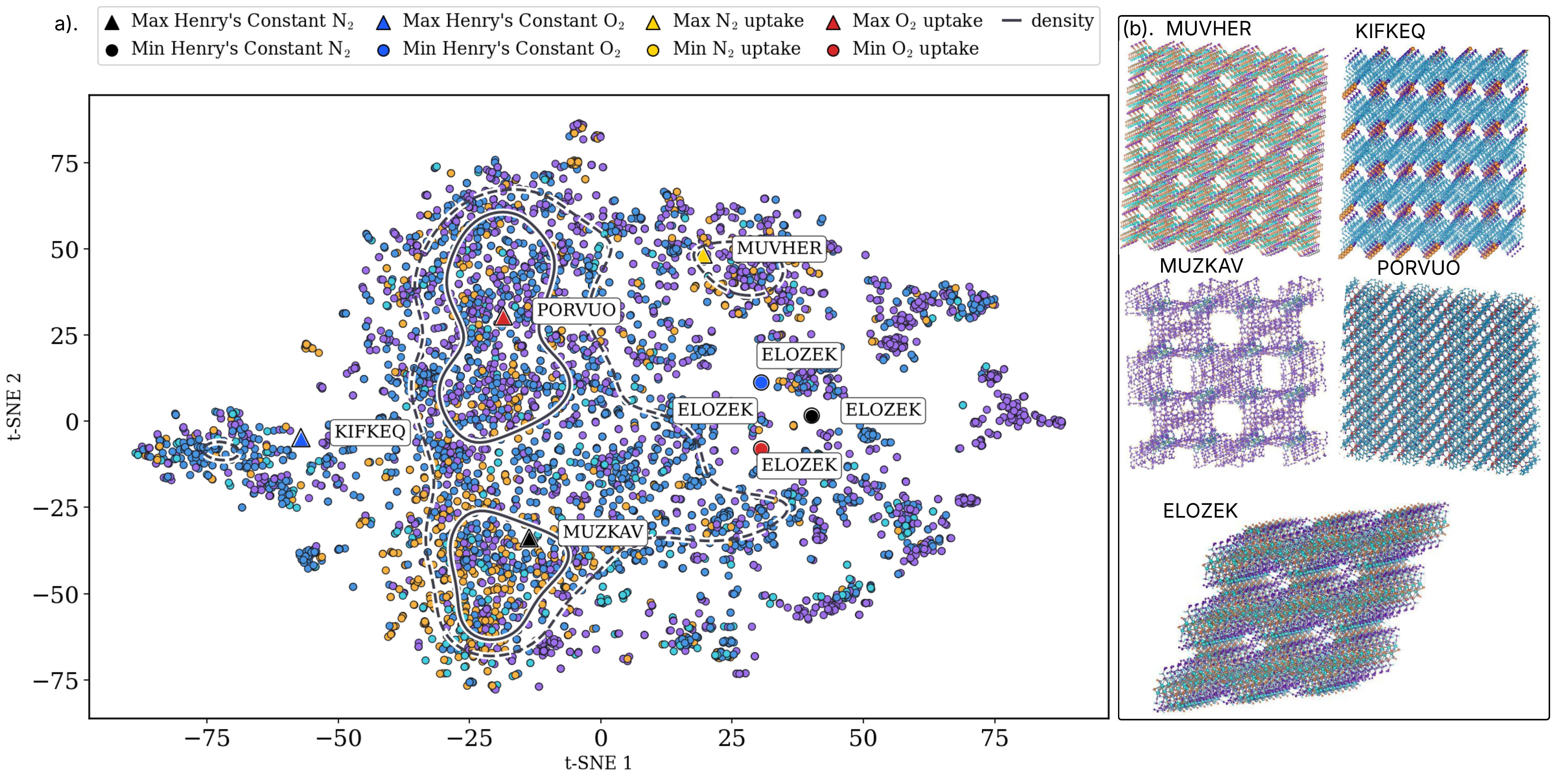}
  \caption{(a). t-SNE of category-specific commutative algebra features. Background points show the dataset; colored markers highlight the maxima and minima for the four properties. We computed two dimensional t-SNE embedding with perplexity equal to $35$, number of iterations equal to $1500$, and random seed equal to $42$ on our standardized $\alpha$ facet and $f$-vector features. Solid contours indicate kernel density level sets at the $85$ percent and 95 percent quartiles. (b) Representative MOFs at property extrema. The selected structures illustrate the diversity of network geometries, pore architectures, and elemental compositions associated with extreme adsorption behaviors. Specifically, MUZKAV exhibits the highest Henry’s constant for N$_2$ ($7.89\times10^{-6}$ mol kg$^{-1}$ Pa$^{-1}$), while KIFKEQ attains the maximum Henry’s constant for O$_2$ ($9.79\times10^{-6}$ mol kg$^{-1}$ Pa$^{-1}$). ELOZEK shows the lowest Henry’s constants for both N$_2$ and O$_2$ ($8.86\times10^{-8}$ mol kg$^{-1}$ Pa$^{-1}$) and also demonstrates the minimum uptake capacities for N$_2$ and O$_2$ (0.0085 mol kg$^{-1}$ and 0.0086 mol kg$^{-1}$, respectively). In contrast, MUVHER and PORVUO exhibit the highest uptake capacities for N$_2$ (0.98 mol kg$^{-1}$) and O$_2$ (1.11 mol kg$^{-1}$), respectively. These structures highlight the range of adsorption responses captured by the CSCA framework. }
  \label{fig:tsne_all_props}
\end{figure}

We partition atoms into element groups $C_a,\ldots,C_h$ and $C_{\mathrm{all}}.$ For each category $C_i\in\{ C_a,\ldots,C_h, C_{\mathrm{all}}\}$, we restrict the previously built $\alpha$-complex to the subcomplex $K(C_i)$. With a radius filtration $\alpha\in[0,12]\,\text{\AA}$ at resolution
$\Delta\alpha=0.1\,\text{\AA}$ (121 thresholds), we record five curves per
category (facet-count in dimensions $0$ and $1$; $f$-vector in dimensions
$1,2,3$), yielding $5\times 121=605$ entries, together with 16 facet summary statistics derived from persistence intervals in dimensions $0$ and $1$. For each dimension, we record the count of intervals, total persistence 
, maximum, mean, median, and the lower and upper (25\%) quartiles of lifetimes . Thus each category contributes $605+16+1=622$ features, and across nine categories $(C_0,\dots,C_7,C_{\mathrm{all}})$ the total is
$9\times 622 = 5{,}598$ features per MOF. The discretized curves are concatenated across all categories to form a single category aware commutative algebra descriptor per MOF. A single 2-D t-SNE of these descriptors is shown in Figure   \ref{fig:tsne_all_props} (a). The figure places MOFs into coherent clusters; we mark property extremes for the four targets with triangles and small circles representing maximum and minimum values of the four properties respectively. The separation of these highlights indicates that the commutative algebra features already encode the key structure property variations relevant to adsorption and transport. This plot illustrates that MOFs with extreme adsorption properties do not scatter randomly but instead fall into recognizable regions of the feature space.

\subsection{Feature Importance}

Facet dimension $1$ and the $f$-vector dimensions $1$ and $2$ provide the dominating signal across categories $C_a, \dots, C_h$ and $C_{\mathrm{all}}$, see Section \ref{supplimentary}.  The majority of the predictive power is enabled by local to medium connectivity, as indicated by the concentration of peaks at small to intermediate filtration scales $\alpha\le5$\,\AA.  Carbon in $C_f$ and hydrogen in $C_e$ rise early in facet dimension $1$ and $f$-vector dimensions $1$ and $2$; oxygen, sulfur, and selenium in $C_h$ give broader profiles; metals in $C_b$ exhibit sharp, confined responses, particularly in $f$-vector dimension $2$; and $f$-vector dimension $3$ only appears as selective, lower magnitude peaks (for instance, in $C_f$ and $C_{\mathrm{all}}$).  These patterns hold for both uptakes and Henry's constants (Figures~\ref{FI_HenryN2}-\ref{FI_O2uptake}).  

Across element based categories $C_a$ to $C_h$ and the pooled set $C_{\mathrm{all}}$, the most informative descriptor families are facet dimension $1$ and $f$-vector dimensions $1$ and $2$, which show dominant peaks at small to intermediate filtration scales ($\alpha \le 5\,\text{\AA}$) in the feature-importance profiles (Figures \ref{FI_HenryN2}--\ref{FI_O2uptake}). The strongest contributions arise from $C_a$ (alkali and alkaline-earth metals), 
$C_g$ (nitrogen--phosphorus), and $C_h$ (oxygen--sulfur--selenium), indicating that short- and medium-range polar or charged environments govern MOF selectivity toward N$_2$ and O$_2$, while framework carbons and nonpolar linkers ($C_f$) primarily shape geometric confinement rather than selective adsorption.

 In $C_a$, facet dimension $1$ and $f$-vector dimension $2$ are significant for all four properties, emphasizing the importance of local carbon framework geometry across tasks.  $C_b$ exhibits abrupt, narrow peaks in $f$-vector dimension $2$, particularly for Henry's constants N$_2$ and O$_2$ in Figure \ref{FI_HenryN2} and Figure \ref{FI_HenryO2}, respectively, indicating hydrogen's effect at highly confined structural scales.  $C_c$ and $C_d$ provide larger, multi-scale contributions, as seen in Figure \ref{FI_N2uptake} and Figure  \ref{FI_HenryO2} for N$_2$ and O$_2$ uptakes respectively, demonstrating that metal-centered connectivity and O-linked nodes impact uptake via longer-range structural organization. Intermediate categories $C_e$ to $C_g$ often include substantial peaks in facet dimension $1$ and $f$-vector dimension $1$, which serve as supportive contributors. Finally, $C_h$ commonly exhibits larger significance profiles, most notably for O$_2$ uptake in Figure \ref{FI_O2uptake}, indicating oxygen's continued engagement in linker node connection, which influences overall transport. Figures \ref{FI_HenryN2} and \ref{FI_HenryO2} for Henry's constants N$_2$ and O$_2$ respectively  show clearly concentrated significance peaks at certain $\alpha$ values, demonstrating sensitivity to micropore environments and first contact adsorption sites. N$_2$ and O$_2$ Uptake in Figure \ref{FI_HenryN2} and  Figure \ref{FI_HenryO2}  respectively show a more dispersed relevance across $\alpha$ and categories, indicating the necessity to reflect multi-scale connectivity and accessible volume that influence loading beyond the initial adsorption event. Despite these variations, the dominance of facet dimension $1$ and $f$-vector dimensions $1$ and $2$ is a continuous cross-property characteristic. These feature analysis demonstrates that descriptor families encoding local-to-medium scale morphology facet dimension $1$, $f$-vector dimension $1$ and $2$ are most predictive, while large $\alpha$ features contribute comparatively little. Different element categories specialize: carbons and hydrogens dominate localized effects, while metals and oxygen extend predictivity over broader scales, especially for uptake.  These data give clear direction for descriptor selection, prioritize facet dimension $1$ and $f$-vector dimensions $1$ and $2$, and imply that models aiming uptake should focus on multi-scale features covering various $\alpha$ ranges.

\section{Method}
\label{sec:Methods}
\subsection{Datasets}\label{dataset}

The dataset comprised of structures from the CoRE MOF 2019 database, which is widely used in comparable work\cite{chen2025category,chung2019advances}. We focus on adsorption properties: Henry coefficients for O$_{2}$ and N$_{2}$, selectivity derived from these coefficients, and uptake capacities. We use the filtering method of Orhan et al. \cite{orhan2021prediction}, removing entries that exceeded property-specific upper bounds and excluding within sigma outliers. As a result, the dataset obtained was more consistent and representative of the property space of interest.
Additionally, upper limit thresholding was employed to reduce bias and boost the robustness of the predictive models\cite{orhan2021prediction}. After cleaning and preprocessing, the dataset was split into training, validation, and test groups using an 80{:}10{:}10 ratio as in \cite{chen2025category}. Table~\ref{tab:2} provides a summary of the final dataset used in this study, along with the properties considered, the corresponding number of samples and the randomized splitting approach employed for model development.

\begin{table}[H]
  \centering
  \caption{Dataset statistics for \(\mathrm{O_2}\) and \(\mathrm{N_2}\) adsorption and transport characteristics in MOFs~\cite{chen2025category}.}
  \begin{tabularx}{\linewidth}{@{}Y Y S[table-format=4.0] c c@{}}
    \toprule
    \textbf{Property} & \textbf{Units} & {\textbf{Samples}} & \textbf{Split} & \textbf{Ratio} \\
    \midrule
    Henry's constant of \(\mathrm{N_2}\) & mol\,kg\(^{-1}\)\,Pa\(^{-1}\) & 4744 & Random & 8:1:1 \\
    Henry's constant of \(\mathrm{O_2}\) & mol\,kg\(^{-1}\)\,Pa\(^{-1}\) & 5036 & Random & 8:1:1 \\
    \(\mathrm{N_2}\) uptake              & mol\,kg\(^{-1}\)             & 5132 & Random & 8:1:1 \\
    \(\mathrm{O_2}\) uptake              & mol\,kg\(^{-1}\)             & 5241 & Random & 8:1:1 \\
    \bottomrule
  \end{tabularx}
  \label{tab:2}
\end{table}

\subsection{Simplicial Complex}
Let \(V=\{1,\dots,n\}\) be a finite set of vertices. A simplex is any finite subset \(\sigma\subseteq V\). A simplicial complex on \(V\) is a family \(K\subseteq 2^V\) satisfying the closure condition
\begin{equation}\label{eq:downward}
  \sigma\in K,\ \tau\subseteq\sigma \;\;\implies\;\; \tau\in K,
\end{equation}
so every face of a simplex is also in \(K\).  
We also require every vertex to appear: \(\{v\}\in K\) for all \(v\in V\).

Elements of \(K\) are called faces. A facet is a face that is not contained in any larger face, and the set of all facets is
\begin{equation}\label{eq:facets}
  \mathcal F(K)\;=\;\{\tau\in K:\ \nexists\,\sigma\in K \text{ with } \tau\subsetneq\sigma\}.
\end{equation}
The dimension of a face is \(\dim\sigma=|\sigma|-1\), and the dimension of the complex is
\begin{equation}\label{eq:dimK}
  \dim K\;=\;\max_{\sigma\in K}\dim\sigma.
\end{equation}
For a vertex subset \(W\subseteq V\), the induced subcomplex is
\begin{equation}\label{eq:induced}
  K_W\;=\;\{\sigma\in K:\ \sigma\subseteq W\}.
\end{equation}

\subsection{Stanley-Reisner theory}
Let $\Bbbk$ be a field and let
\[
  S=\Bbbk[x_1,\dots,x_n], \qquad \deg x_i=1,
\]
be the polynomial ring whose variables $x_i$ correspond to the vertices $i\in V$. To each face \(\sigma\subseteq V\) associate the squarefree monomial
\[
  x^\sigma=\prod_{i\in\sigma}x_i, \qquad x^\varnothing=1
\]
so that each subset of vertices is encoded by a distinct monomial. The Stanley--Reisner ideal of \(K\) records the nonfaces:
\begin{equation}\label{eq:sr-ideal}
  I(K)\;=\;\langle\,x^\sigma:\ \sigma\subseteq V,\ \sigma\notin K\,\rangle \subseteq S,
\end{equation}
and the corresponding Stanley--Reisner ring is
\begin{equation}\label{eq:sr-ring}
  \Bbbk[K]\;=\;S/I(K).
\end{equation}
Thus, monomials corresponding to actual faces remain nonzero in $\Bbbk[K]$, while those corresponding to nonfaces vanish. For each facet $\tau\in\mathcal F(K)$, the facet prime is
\begin{equation}\label{eq:facet-prime}
  P_\tau=(x_i:\ i\notin\tau)\ \subset S.
\end{equation}
These are the minimal prime ideals of \(I(K)\), and one has the irredundant decomposition
\begin{equation}\label{eq:decomp}
  I(K)\;=\;\bigcap_{\tau\in\mathcal F(K)}P_\tau.
\end{equation}

A minimal graded free resolution of $\Bbbk[K]$ over the polynomial ring $S=\Bbbk[x_1,\dots,x_n]$
is an exact sequence of free $S$-modules
\begin{equation}\label{eq:resolution}
  \cdots \;\longrightarrow\;
  \bigoplus_{j} S(-j)^{\,\beta_{i,j}}
  \;\longrightarrow\;
  \cdots
  \;\longrightarrow\;
  \bigoplus_{j} S(-j)^{\,\beta_{0,j}}
  \;\longrightarrow\;
  \Bbbk[K]
  \;\longrightarrow\; 0.
\end{equation}
Each term $S(-j)$ represents a graded shift of $S$ by degree $j$,  
and the integers $\beta_{i,j}$ are the graded Betti numbers that record how many generators or relations occur at each homological degree $i$ and internal degree $j$.  
They can be computed algebraically as
\begin{equation}\label{eq:betti-def}
  \beta_{i,j}
  \;=\;
  \dim_{\Bbbk}
  \big(\operatorname{Tor}^{S}_{i}(\Bbbk[K],\Bbbk)\big)_{j},
\end{equation}
where $\operatorname{Tor}^{S}_{i}(\Bbbk[K],\Bbbk)$ measures the dependencies among generators at level $i$,  
and the subscript $j$ extracts the component of total degree $j$.

Hochster’s formula connects them to simplicial homology:
\begin{equation}\label{eq:hochster}
  \beta_{i,i+j}(\Bbbk[K]) \;=\;
  \sum_{\substack{W\subseteq V\\ |W|=i+j}}
  \dim_{\Bbbk}\,\widetilde H_{j-1}(K_W;\Bbbk).
\end{equation}
Here $K_W$ is the subcomplex induced by vertex subset $W$, and
$\widetilde H_{j-1}(K_W;\Bbbk)$ denotes the reduced homology group in dimension $j\!-\!1$ with coefficients in $\Bbbk$.  
Its dimension counts the number of independent $(j\!-\!1)$-dimensional holes in $K_W$.

If \(\dim K=d-1\), the Hilbert series of \(\Bbbk[K]\) has the form
\begin{equation}\label{eq:hilbert}
  H_{K}(t)\;=\;\frac{h_0+h_1t+\cdots+h_d t^{d}}{(1-t)^d}.
\end{equation}
The coefficients \(h(K)=(h_0,\dots,h_d)\) define the $h$-vector.  
The $f$-vector counts faces by dimension:
\begin{equation}\label{eq:fvector}
  \mathbf f(K)=(f_{-1},f_0,\dots,f_{d-1}),\quad 
  f_i=\#\{\sigma\in K:\dim\sigma=i\},\quad f_{-1}=1.
\end{equation}
They are related by
\begin{equation}\label{eq:h-f}
  \sum_{j=0}^d h_j t^j \;=\; \sum_{j=0}^d f_{j-1} t^j(1-t)^{d-j}.
\end{equation}

\subsection{Filtration}
The \v{C}ech complex, the $\alpha$-complex, and the Vietoris Rips complex \cite{ghrist2014elementary, masood2020parallel, chambers2010vietoris} are the three filtrations that are often employed in topological data analysis. All three yield nested sequences of simplicial complexes that enlarge as the scale parameter increases.  
We use the $\alpha$-complex in this study, which is geometrically appropriate for atomic coordinates in periodic MOFs. A simplex (edge, triangle, or tetrahedron) is added once the radius of its lowest empty circumsphere is at most $\alpha$. It is built using the Delaunay triangulation. Figure \ref{filtration} shows an illustration of a filtration. 
As $\alpha$ grows, the complexes expand by subset inclusion,
\begin{equation}\label{eq:filtration}
  K^{\le \alpha}\subseteq K^{\le \alpha'} \quad (\alpha\le\alpha'), 
  \qquad \bigcup_{\alpha\ge0}K^{\le\alpha}=K.
\end{equation}

The birth of a face $\sigma$ is the smallest scale where it appears:
\begin{equation}\label{eq:birth}
  b(\sigma)=\inf\{\alpha:\ \sigma\in K^{\le \alpha}\}.
\end{equation}
The facet intervals in dimension $i$ record the persistence of facets of dimension $i$:
\begin{equation}\label{eq:facet-interval}
  \mathsf{FI}^{(i)}(K)\;=\;\{[\hat b(\tau),\hat d(\tau)):\ \tau \text{ an $i$-facet at some scale}\}.
\end{equation}
Here $\hat b(\tau)$ and $\hat d(\tau)$ denote the birth and death scales of the facet $\tau$, respectively.  
The birth scale $\hat b(\tau)$ is the smallest value of $\alpha$ for which $\tau$ appears in the filtration $K^{\le\alpha}$,  
while the death scale $\hat d(\tau)$ is the smallest value of $\alpha$ at which $\tau$ ceases to be a facet,  
that is, when it becomes a face of a higher-dimensional simplex. Therefore, each interval $[\hat b(\tau), \hat d(\tau))$ represents how long the facet $\tau$ persists during the filtration.

The combinatorial growth of the complex is summarized by the $f$-vectors,
\begin{equation}\label{eq:f-curves}
  f_i(\alpha)\;=\;\#\{\sigma\in K^{\le \alpha}:\ \dim\sigma=i\},\qquad i\ge0,
\end{equation}
which track how many vertices, edges, triangles, and tetrahedra are present at each scale.

\subsection{Category-Restricted Complexes}
Let $\mathcal G=\{a,b,c,d,e,f,g,h,\mathrm{all}\}$ denote the element categories.  
For each $i\in\mathcal G$, define
\begin{equation}\label{eq:cat-complex}
  K_i=K(C_i),
\end{equation}
the subcomplex induced on vertices of category $C_i$ ($K_{\mathrm{all}}$ is the full complex).  
Each $K_i$ inherits the $\alpha$-complex filtration \eqref{eq:filtration}, so that
\[
  K_i^{\le \alpha}\subseteq K_i^{\le \alpha'} \quad (\alpha\le\alpha').
\]
For every category and dimension we record:
\begin{align}
  F^{(i)}_j(\alpha) &= \{\sigma\in K_i^{\le \alpha}:\ \dim\sigma=j\}, \label{eq:Fij}\\
  f^{(i)}_j(\alpha) &= |F^{(i)}_j(\alpha)|, \label{eq:fij}\\
  \mathrm{Facet}^{(i)}_j(\alpha) &= \{\tau\in K_i^{\le \alpha}:\ \tau \text{ a $j$-facet}\}. \label{eq:facetij}
\end{align}
Facet intervals for category $i$ are
\begin{equation}\label{eq:fi-cat}
  \mathsf{FI}^{(j)}(K_i)\;=\;\{[\hat b(\tau),\hat d(\tau)):\ \tau\in \mathrm{Facet}^{(i)}_j(\alpha)\}.
\end{equation}
The $f$-vector curves are
\begin{equation}\label{eq:fcurves-cat}
  \mathbf f^{(i)}_{1,2,3}(\alpha)=\big(f^{(i)}_1(\alpha),\,f^{(i)}_2(\alpha),\,f^{(i)}_3(\alpha)\big),
\end{equation}
which describe the evolution of edges, triangles, and tetrahedra within each element category.

\subsection{Feature Vector}
The final feature vector concatenates, over all category features,  
statistical summaries of facet intervals $\mathsf{FI}^{(j)}(K_i)$ for $j=0,1$,  
together with sampled values of the $f$-vector curves \eqref{eq:fcurves-cat}.  
This combines persistence based information from facets with combinatorial counts from $f$-vectors,  
providing a joint representation of how structural connectivity evolves across scales and categories.

\begin{figure}[H]
    \centering
    \includegraphics[width=0.8\linewidth]{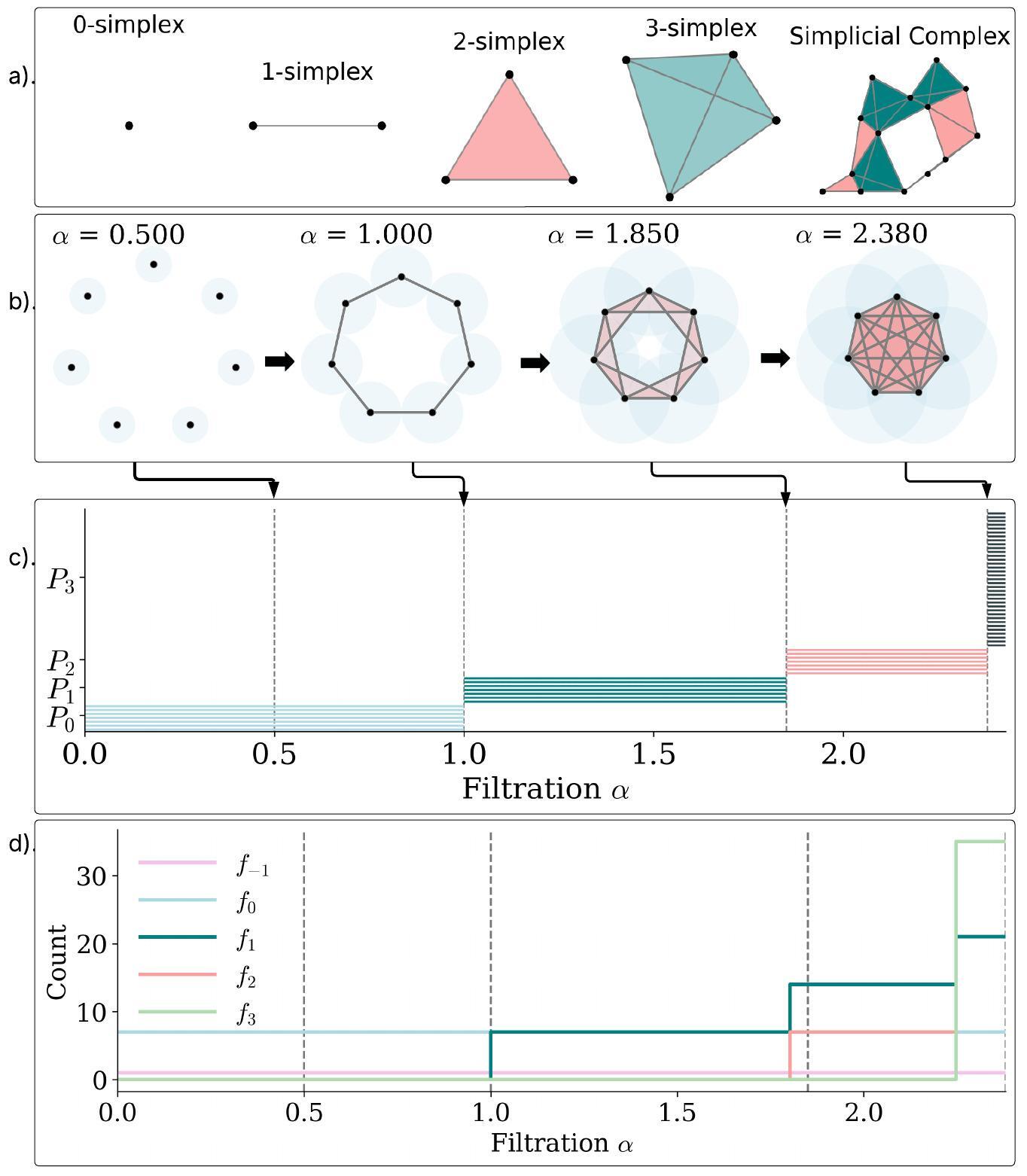}
  \caption{Simplicial filtration and summary statistics. a) The representation of simplices in dimensions $0,1,2$ and $3$: (b) filtration of an $\alpha$-complex on a $2$ dimension heptagon random point cloud, $\alpha$-complex snapshots for $\alpha\in\{0.50,\,1.00,\,0.98,\,1.85,\,2.38\}$; previously present simplices are faded, newly added ones are highlighted; translucent $\alpha$-balls show the scale: (c) Summary plots: facet barcodes by dimension ($P_0, P_2, P_3$): (d) $f$-vector curves $f_k(\alpha)$ for $k=-1,0,1,2,3$.}
  \label{filtration}
\end{figure}

\subsection{Category-Specific Commutative Algebra Embedding}
Graph-based models are frequently used in traditional representations of MOF topology, in which bonds are represented as edges and atoms as nodes\cite{chen2025category}. This model is useful for capturing pairwise interactions, but it is unable to represent higher-order structural connections that are essential for stability, adsorption, and diffusion\cite{chen2025category}. A particular type of simplicial complex that offers a more sophisticated algebraic topological framework is alpha complexes, which we use to get around this restriction. Due to their derivation from the Delaunay triangulation, alpha complexes maintain geometric accuracy while drastically lowering duplication, in contrast to general Vietoris Rips complexes \cite{carlsson2024computing}. Our method first scales MOF structures into a homogeneous supercell of $64$\AA $\times 64$\AA $\times 64$\AA, ensuring uniformity across all datasets. Following that, atoms are categorized into several groups based on their structural properties and chemical similarities, as shown in Table~\ref{tab:3}. In order to summarize higher order simplices, we compute $f$-vectors in dimensions $1, 2$, and $3$ and extract facet intervals of dimensions $0$ and $1$ from these complexes.  We obtain summary statistics of the corresponding filtration values, including the mean, variance, minimum, and maximum, for each category and simplex dimension in order to make the representation robust and comparable. The analysis is limited to physically relevant scales by truncating the filtration at a maximum radius  $\alpha_{\max}=12$\AA. Features for machine learning and analysis are obtained by concatenating the interval features, $f$-vectors, and summary statistics over all categories.

Figure \ref{filtration} shows a filtration built by a simple ball rule:
at each scale $\alpha$ we draw a ball of radius $\alpha$ around every point.
A vertex is present for all $\alpha$; an edge appears when the two
balls intersect; a triangle appears when all three pairwise edges are
present (equivalently, the three balls have a common intersection); and a
tetrahedron appears when all six edges among four vertices are present. Thus, at $\alpha=0.50$ only vertices are present; at $\alpha=1.00$ the outer-ring edges appear; at $\alpha=1.85$ the
first triangles appear; and by $\alpha=2.38$ tetrahedra also appear. For each facet $\sigma$ we define its birth $b(\sigma)$ as the smallest $\alpha$ at which it first exists under this rule, and its death $d(\sigma)$ as the smallest $\alpha$ at which it first becomes a face of a higher-dimensional facet.
The barcode in panel (b) of Figure \ref{filtration} draws one $[b(\sigma),\,d(\sigma)\,)$ per facet, grouped by dimension
$\mathcal{P}_k$, where $\mathcal{P}_0$ denotes vertices (dimension 0),
$\mathcal{P}_1$ edges (dimension 1), $\mathcal{P}_2$ triangles (dimension 2), and
$\mathcal{P}_3$ tetrahedra (dimension 3). Accordingly, panel (b) shows 7 vertex bars in $[0,1.00)$, $7$ edge bars in $[1.00,1.85)$, 7 triangle bars in $[1.85,2.38)$,
and 35 tetrahedron bars at $[2.38,\infty)$. Panel (c) tracks how the number of active facets evolves with $\alpha$ which is represented by the $f$-vector curves $f_k(\alpha)$ where $k$ is the dimension. We plot $f_k(\alpha)=\#\{\text{$k$ facets alive at }\alpha\}$, so curves step up at births, levels when the facets persists and step down at deaths. At the marked scales the values (in order $(f_{-1},f_{0},f_{2},f_{3})$) are:
$0.50:(1,7,0,0)$, $1.00:(1,7,0,0)$, $1.85:(1,7,7,0)$, $2.38:(1,7,35,35)$.

\begin{table}[H]
  \centering
  \caption{Element categories $\mathcal{C}_a,\dots,\mathcal{C}_h,\ \mathcal{C}_{\mathrm{all}}$ used for category-specific topological modeling of MOFs.}
  \label{tab:elem-cats}
  \begin{tabularx}{\linewidth}{@{}Y c Y@{}}
    \toprule
    \textbf{Element category} & \textbf{Notation} & \textbf{Elements included} \\
    \midrule
    Alkali metals, alkaline earth metals, and post-transition metals 
      & $C_a$ 
      & Li, Na, K, Rb, Cs; Be, Mg, Ca, Sr, Ba; Al, Ga, In, Sn, Pb, Bi \\
    Transition metals, lanthanides, actinides 
      & $\mathcal{C}_b$ 
      & Sc, Ti, V, Cr, Mn, Fe, Co, Ni, Cu, Zn; Y, Zr, Nb, Mo, Ru, Rh, Pd, Ag, Cd; Hf, Ta, W, Re, Os, Ir, Pt, Au, Hg; La, Ce, Pr, Nd, Sm, Eu, Gd, Tb, Dy, Ho, Er, Tm, Yb, Lu; Th, U \\
    Metalloids 
      & $C_c$ 
      & B, Si, Ge, As, Sb, Te \\
    Halogens 
      & $C_d$ 
      & F, Cl, Br, I \\
    Hydrogen 
      & $C_e$ 
      & H \\
    Carbon 
      & $C_f$ 
      & C \\
    Nitrogen, phosphorus 
      & $C_g$ 
      & N, P \\
    Oxygen, sulfur, selenium 
      & $C_h$ 
      & O, S, Se \\
    All elements (union of $C_a$, \dots, $C_h$) 
      & $C_{\mathrm{all}}$ 
      & All of the above \\
    \bottomrule
  \end{tabularx}
  \label{tab:3}
\end{table}

\subsection{Model Prediction}
In this work, we use a gradient boosted decision tree (GBDT) regressor to predict the target properties from our category-specific commutative algebra (CSCA) feature vectors.  Gradient boosting transforms poor learners into strong predictors by creating an additive ensemble of shallow trees that are each fitted to the current model's residuals under a squared error target \cite{chen2025category}.

\begin{table}[H]
  \centering
  \caption{Gradient Boosting Regressor hyperparameters.}
  \label{tab:gbr-params}
  \begin{tabular}{@{}ll@{}}
    \toprule
    \textbf{Hyperparameter}        & \textbf{Value} \\ 
    \midrule
    Max tree depth                 & 7 \\
    Max features per split         & $\sqrt{p}$ \\
    Min samples per leaf           & 1 \\
    Min samples to split           & 2 \\
    Learning rate                  & 0.005 \\
    Subsample                      & 0.5 \\
    Number of estimators           & \num{10000} \\
    \bottomrule
  \end{tabular}
  \vspace{0.25em}
  \label{tab:4}

  \footnotesize\emph{Note.} $p$ denotes the number of input features.
\end{table}

We optimized the squared error loss function by implementing the gradient boosting regressor from Scikit-learn\cite{pedregosa2011scikit}. The parameters of our model are displayed in Table \ref{tab:4}. The targets were standardized, and all inputs were normalized. The datasets are divided as shown in Table \ref{tab:2}; the validation split is not used for model selection because hyperparameters are fixed. We perform ten random splits and, within each split, train ten models with different seeds (100 training runs per dataset) to reduce variance from random partitioning and initialization. We provide a solid, comparable evaluation to previous work by reporting RMSE, MAE, and the  coefficient of determination, represented as $R^2$, as an average across runs on the 10\% test set and, for completeness, on the pooled 20\% (validation and test).

\section{Conclusion}
\label{sec:conclusion}

 Metal-organic frameworks (MOFs) are highly diverse and adaptable due to the mix of inorganic nodes and organic linkers, leading to a large surface area, tunable porosity, and a wide range of chemical compositions. However, identifying the structure-property relationships that govern adsorption and selectivity remains a significant challenge, especially when analyzing thousands of frameworks with varying compositions and architectures. Traditional experimental and computational techniques have limited capacity to generalize across the vast chemical and structural environment and grasp the intricate links between geometry and function.

This study introduced commutative algebra modeling and prediction, the first of its kind, to materials science. We develop category-specific commutative algebra (CSCA) machine learning to predict gas adsorption properties in MOFs. CSCA represents each framework using persistent facet ideals and f-vectors, calculated for chemically relevant atom groups within specific categories. This algebraic representation converts geometric connectivity into descriptors that are both rigorous and interpretable, thereby linking combinatorial algebra with material structure. The findings indicate that descriptors derived from low-dimensional facet ideals; dimensions $1$ and $2$ and f-vector components; dimensions $1, 2,$ and $3$ consistently provide the most informative predictions for all four properties examined. In conclusion, CSCA machine learning furnishes a category-aware, algebraic representation of MOFs that yields accurate and interpretable predictors of gas adsorption. By computing facet-ideal statistics and $f$-vectors within and across chemically meaningful categories, the method links local composition with global connectivity in a way that is both rigorous and model-agnostic. The proposed commutative algebra approach has potential to tackle a wide variety of other challenges in materials science.  

\section*{Data And Code Availability}

All data sets originate from CoRE MOFs 2019 structures\cite{chung2019advances}; Property calculations for each dataset follow the approach outlined by Orhan et al\cite{orhan2021prediction}. The code and data used in this study are provided at \url{https://github.com/CSCA-MOFs/MOF-CSCA}
, with implementation details described in \cite{grayson2002macaulay2}.

\section{Acknowledgments}
This work was supported in part by NIH grant R35GM148196, National Science Foundation grant DMS2052983,  Michigan State University Research Foundation, and  Bristol-Myers Squibb 65109.  
C.-L.C. gratefully acknowledges financial support from the Defense Threat Reduction Agency (Project CB11141), and the Department of Energy (DOE), Office of Science, Office of Basic Energy Sciences (BES) under an award FWP 80124 at Pacific Northwest National Laboratory (PNNL). PNNL is a multiprogram national laboratory operated for the Department of Energy by Battelle under Contract DE-AC05-76RL01830.
C.-L. C. and G.-W. W would like to also acknowledge financial support from DOE, Office of Science, BES under an award FWP 84274 for the development of machine learning models that are suitable for predicting the binding of clusters with substrate.

\clearpage
\setcounter{page}{2}   

\appendix
\addcontentsline{toc}{section}{Supplementary Material}

\renewcommand{\thesection}{S\arabic{section}}
\renewcommand{\thefigure}{S\arabic{figure}}
\renewcommand{\thetable}{S\arabic{table}}
\setcounter{section}{0}\setcounter{figure}{0}\setcounter{table}{0}

\section{Supplementary materials}
\label{supplimentary}
\subsection{Performance Measures}
We reported Root Mean Square Error (RMSE), Mean Absolute Error (MAE), and the coefficient of determination ($r^2$). Lower RMSE, lower MAE and higher $R^2$ indicated better performance. 
We used RMSE to capture overall prediction error with sensitivity to large deviations. 
We included MAE to provide a more interpretable measure of average error. 
We also reported $r^2$ to evaluate how much variance in the data was explained by our models.

\begin{equation}
\mathrm{RMSE} = \sqrt{\frac{1}{n}\sum_{i=1}^n(\hat{y}_i - y_i)^2},
\end{equation}

\begin{equation}
\mathrm{MAE} = \frac{1}{n}\sum_{i=1}^n \lvert \hat{y}_i - y_i \rvert,
\end{equation}

\begin{equation}
R^2 = 1 - \frac{\sum_{i=1}^n (y_i - \hat{y}_i)^2}{\sum_{i=1}^n (y_i - \bar{y})^2},
\end{equation}

where \(n\) is the number of samples, \(y_i\) is the true value of sample \(i\), 
\(\hat{y}_i\) is the predicted value for sample \(i\), and \(\bar{y}\) is the mean of the true values.

\subsection{Benchmark Dataset Alignment}
In this study we evaluated our model on four benchmark datasets describing the adsorption properties of MOFs: Henry's constants and uptake capacities for N$_2$ and O$_2$.  The four datasets were adapted from the collection introduced by Orhan et al.~\cite{orhan2021prediction}, available through the public repository 
(\url{https://github.com/ibarishorhan/MOF-O2N2/tree/main/mofScripts}). Table \ref{tab:datasets_comparison} compares the size of our dataset with those of the three benchmark datasets. The Descriptor-based~\cite{orhan2021prediction}, MOF Transformer~\cite{kang2023multi}, and PM Transformer~\cite{park2023enhancing} models all rely on datasets from the same source, but the precise dataset details they used were not clearly described.We gathered and summarized the dataset sizes for each method in Table~\ref{tab:datasets_comparison} for fairness and transparency.
\begin{table}[H]
  \centering
  \caption{Comparison of dataset sizes used in this study and previous MOF models.}
  \begin{tabularx}{\linewidth}{@{}l *{5}{>{\centering\arraybackslash}X}@{}}
    \toprule
    \text{Property} & \text{CSCA} & Descriptor-based\cite{orhan2021prediction}
 & MOF Transformer\cite{kang2023multi} 
 & PM Transformer\cite{park2023enhancing} \\
    \midrule
    Henry's constant \(\mathrm{N_2}\) & 4744 & 4755 & - & - \\
    Henry's constant \(\mathrm{O_2}\) & 5036 & 5045 & - & - \\
    \(\mathrm{N_2}\) uptake  & 5132 & 5158 & 5286 & 5286 \\
    \(\mathrm{O_2}\) uptake  & 5241 & 5259 & 5286 & 5286 \\
    \bottomrule
\end{tabularx}
  \label{tab:datasets_comparison}
\end{table}

\subsection{Feature Importance}
\begin{figure}[H]
  \centering
  \includegraphics[width=1\linewidth]{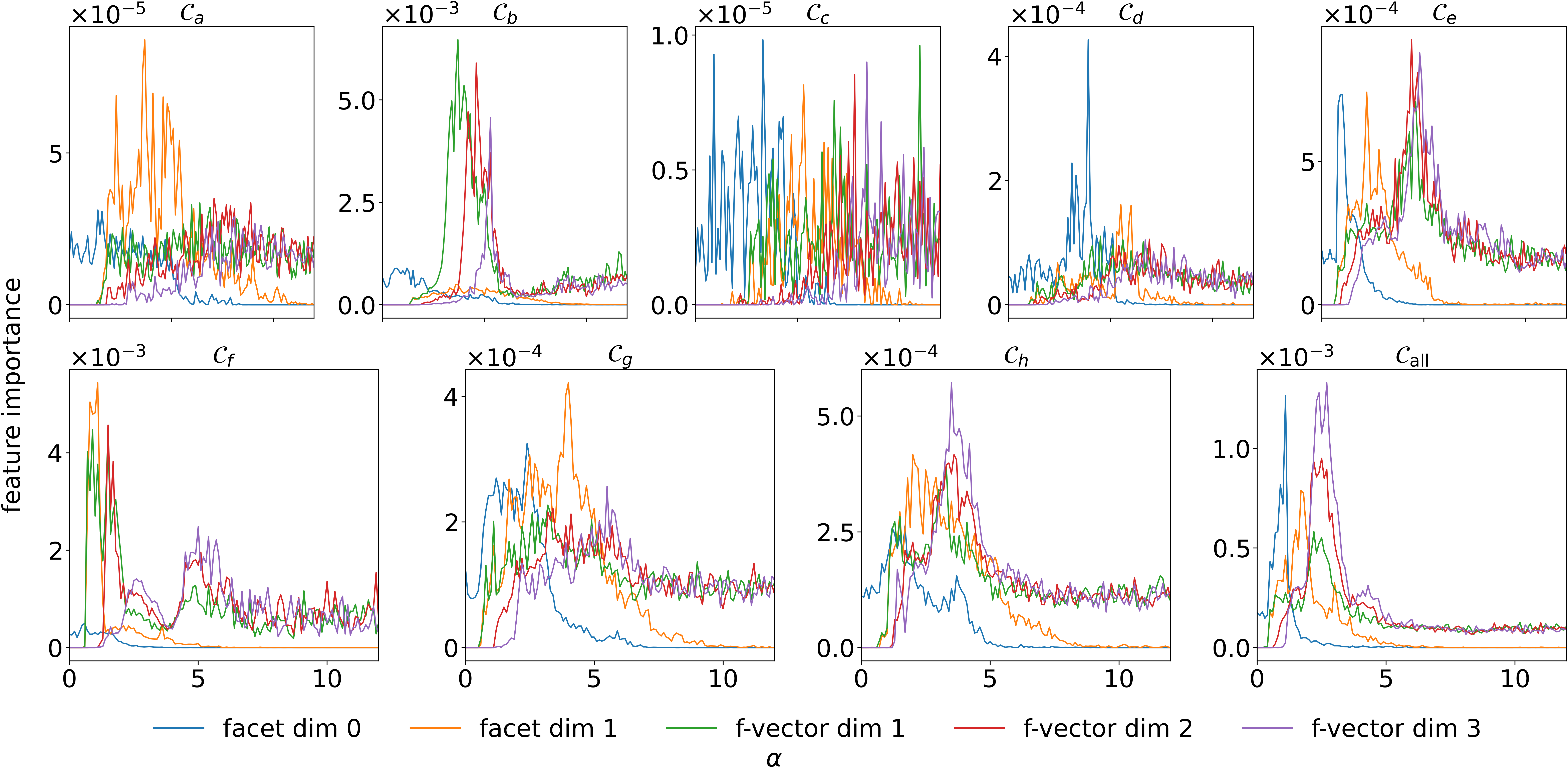}
  \caption{Distribution of feature importance across element-based categories $C_a,\dots,C_h$ and $C_{\mathrm{all}}$ for Henry’s constant of O$_2$. Each panel shows random forest importance as a function of the filtration threshold $\alpha$, with curves corresponding to facet dimensions $0$ and $1$ and $f$-vector dimensions $1$, $2$, and $3$}
  \label{FI_HenryN2}
\end{figure}

\begin{figure}[H]
  \centering
  \includegraphics[width=1\linewidth]{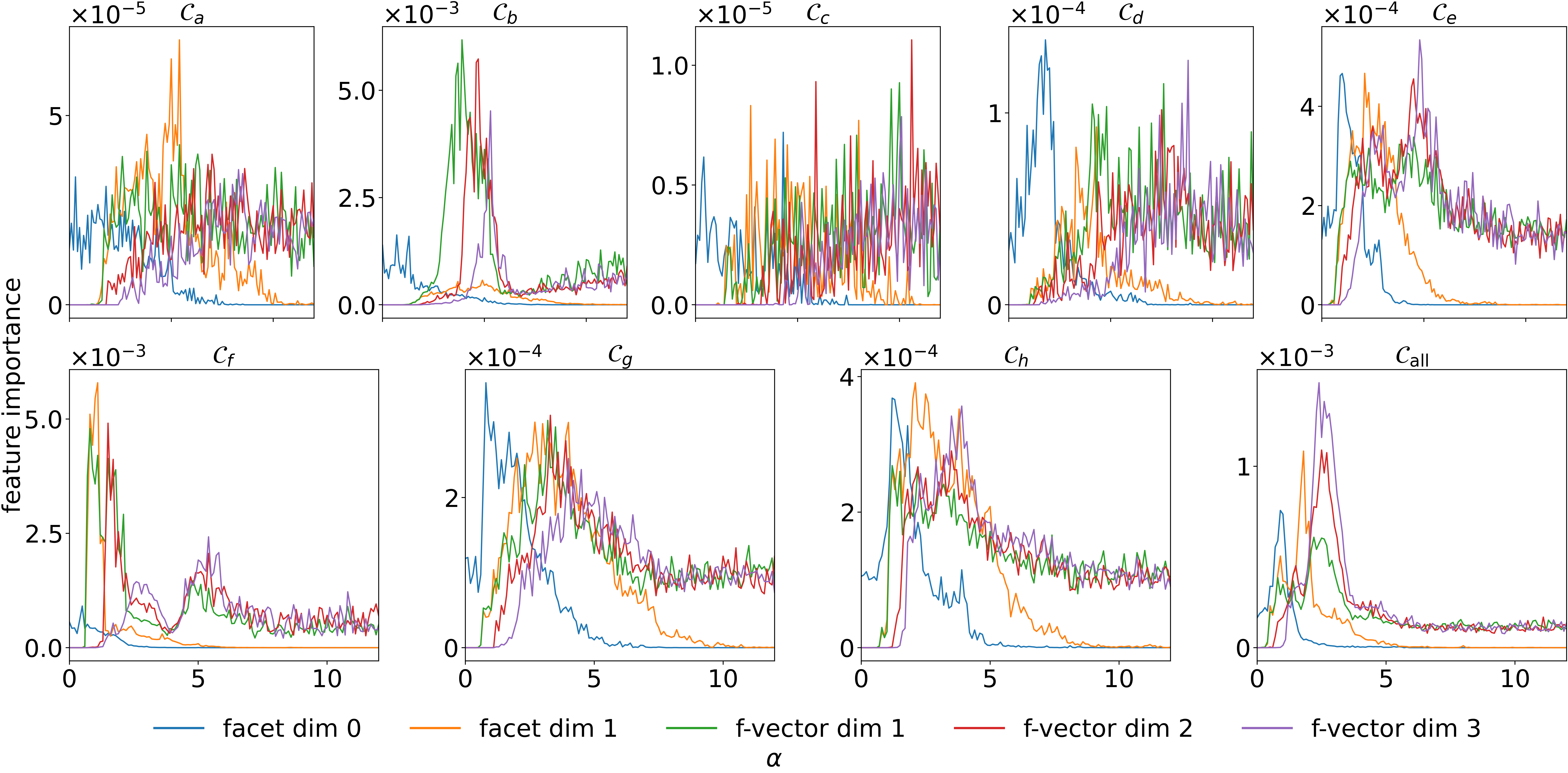}
  \caption{Distribution of feature importance across element-based categories $C_a,\dots,C_h$ and $C_{\mathrm{all}}$ for N$_2$ uptake capacity. Each panel shows random forest importance as a function of the filtration threshold $\alpha$, with curves corresponding to facet dimensions $0$ and $1$ and $f$-vector dimensions $1$, $2$, and $3$}
  \label{FI_N2uptake}
\end{figure}

\begin{figure}[H]
  \centering
  \includegraphics[width=1\linewidth]{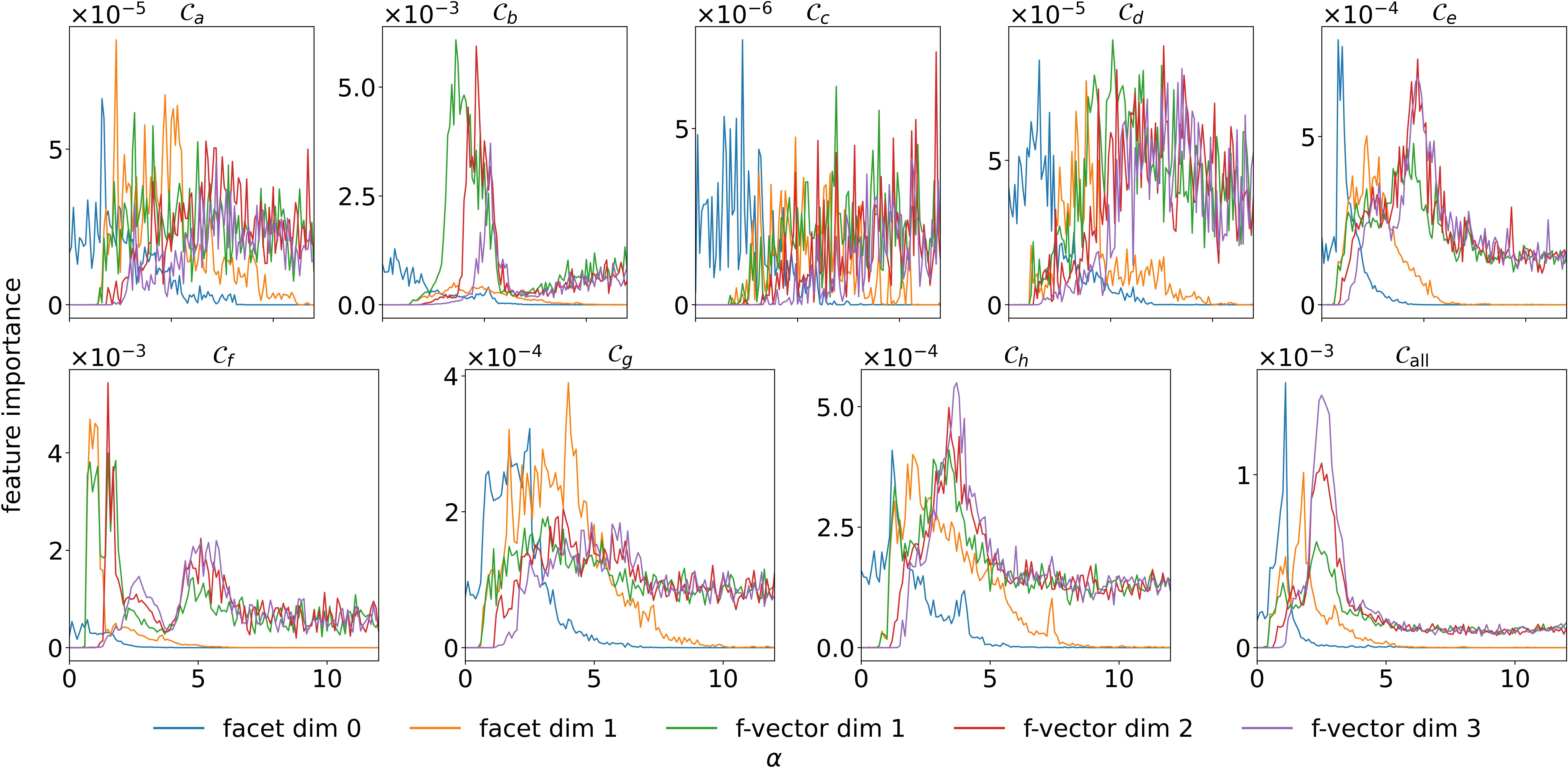}
  \caption{Distribution of feature importance across element-based categories $C_a,\dots,C_h$ and $C_{\mathrm{all}}$ for O$_2$ uptake capacity. Each panel shows random forest importance as a function of the filtration threshold $\alpha$, with curves corresponding to facet dimensions $0$ and $1$ and $f$-vector dimensions $1$, $2$, and $3$.}
  \label{FI_O2uptake}
\end{figure}

The distribution of feature importance across the eight element-based categories $C_a,\dots,C_h$ shows how each descriptor family contributes to four adsorption characteristics, as illustrated in Figures ~\ref{FI_HenryN2}-\ref{FI_O2uptake}. For a particular category, each tiny panel displays the aggregated random forest importance as a function of the filtering threshold $\alpha$, with distinct curves for facet dimension $0$, facet dimension $1$, and $f$-vector dimensions $ 1$, $2$, and $ 3$. Figures ~\ref{FI_HenryN2}-\ref{FI_O2uptake} represent Henry's constants N$_2$, Henry's constant O$_2$, N$_2$ uptake, and O$_2$ uptake, respectively, while each subplot represents categories $C_a$ through $C_h$ and $C_{\mathrm{all}}$. For each property, we trained a \texttt{RandomForestRegressor} with $100$ trees, \texttt{max\_features} set to \texttt{sqrt}, and a fixed seed for consistency.  We utilized a 3-fold cross-validation repeated 5 times, with feature importances averaged over all folds and repeats.

\begin{figure}[H]
  \centering
  \includegraphics[width=1\linewidth]{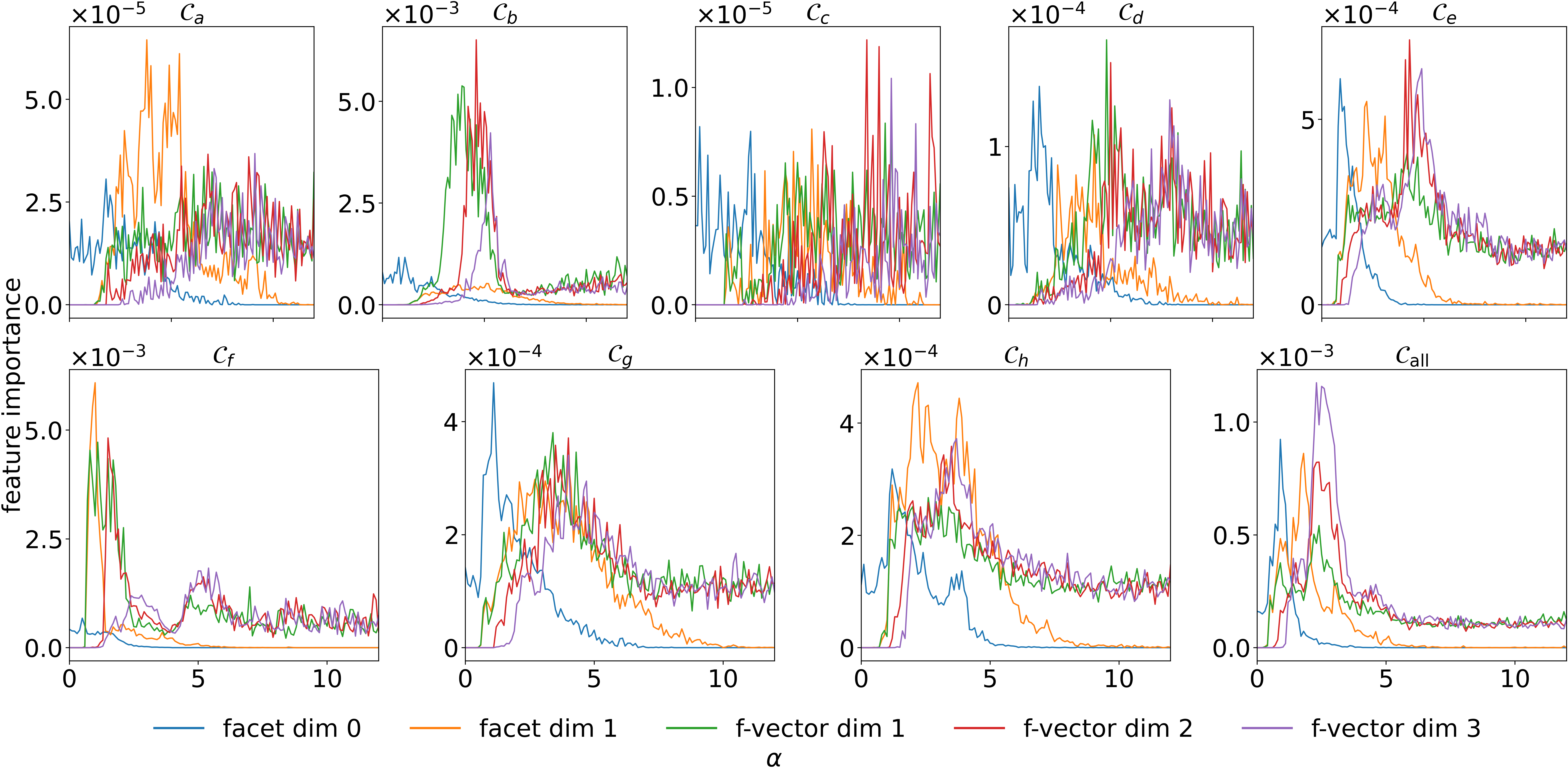}
  \caption{Distribution of feature importance across element-based categories $C_a,\dots,C_h$ and $C_{\mathrm{all}}$ for Henry’s constant of N$_2$. Each panel shows random forest importance as a function of the filtration threshold $\alpha$, with curves corresponding to facet dimensions $0$ and $1$ and $f$-vector dimensions $1$, $2$, and $3$.}
  \label{FI_HenryO2}
\end{figure}

\bibliographystyle{unsrt}
\bibliography{references}
\end{document}